\documentclass{ws-ijmpe}
\usepackage[super,compress]{cite}

\begin{document}

\markboth{M. Thoennessen}{2017 Update of the Discoveries of Isotopes}

\catchline{}{}{}{}{}

\title{2017 UPDATE OF THE DISCOVERIES OF NUCLIDES}

\author{\footnotesize M. THOENNESSEN}

\address{National Superconducting Cyclotron Laboratory and 
Department of Physics \& Astronomy \\
Michigan State University 
East Lansing, Michigan 48824 and\\
American Physical Society, Ridge, New York 11961\\
thoennessen@nscl.msu.edu}

\maketitle

\begin{history}
\received{Day Month Year}
\revised{Day Month Year}
\end{history}

\begin{abstract}
The 2017 update of the discovery of nuclide project is presented. 34 new nuclides were observed for the first time in 2017. However, the assignment of six previously identified nuclides had to be retracted.
\end{abstract}


\ccode{PACS numbers: 21.10.-k, 29.87.+g}


\section{Introduction}

This is the fifth update of the isotope discovery project which was originally published in a series of papers in Atomic Data and Nuclear Data Tables from 2009 through 2013 (see for example the first \cite{2009Gin01} and last \cite{2013Fry01} papers). Two summary papers were published in 2012 and 2013 in Nuclear Physics News \cite{2012Tho03} and Reports on Progress in Physics \cite{2013Tho02}, respectively, followed by annual updates in 2014 \cite{2014Tho01}, 2015 \cite{2015Tho01}, 2016 \cite{2016Tho02} and 2017 \cite{2017Tho01}. In  2016 a description of the discoveries from a historical perspective was published in the book ``The Discovery of Isotopes -- A complete Compilation''  \cite{2016Tho01}.

\section{New discoveries in 2017}
\label{New2017}

In 2017, the discoveries of 34 new nuclides were reported in refereed journals. As in the previous two years the majority were discovered in fragmentation reactions at RIKEN in Japan. The three neutron-rich cesium isotopes $^{149-151}$Cs had been included in previous compilations but their assignments were retracted (see below in section \ref{Changes2017}). Three $\alpha$-emitting transuranium nuclides were identified for the first time in fusion-evaporation reactions at Jyv\"askyl\"a in Finland and Lanzhou in China. In addition, one nuclide was observed in reactions from secondary beams at Michigan State University in the U.S. Table \ref{2017Isotopes} lists details of the discovery including the production method. 

\begin{table}[pt]
\tbl{New nuclides reported in 2017. The nuclides are listed with the first author, submission date, and reference of the publication, the laboratory where the experiment was performed, and the production method (PF = projectile fragmentation, FE = fusion evaporation, SB = secondary beams)\label{2017Isotopes}.}
{\begin{tabular}{@{}llrclc@{}} \toprule 
Nuclide(s) & First Author & Subm. Date & Ref. & Laboratory & Type \\ \colrule
$^{73}$Mn, $^{76}$Fe, $^{78}$Co,  	&  T. Sumikama & 7/15/2016 &\refcite{2017Sum01} & RIKEN & PF \\
$^{81}$Ni, $^{82}$Ni, $^{83}$Cu 	& & & & & \\
$^{149}$Cs, $^{150}$Cs, $^{151}$Cs, $^{153}$Ba, 	& J. Wu & 9/29/2016 &\refcite{2017Wu01} & RIKEN & PF \\
$^{154}$Ba, $^{154}$La, $^{155}$La, $^{156}$La, 	& & & & & \\
$^{156}$Ce, $^{157}$Ce, $^{158}$Ce, $^{156}$Pr,& & & & & \\
$^{157}$Pr, $^{158}$Pr, $^{159}$Pr, $^{160}$Pr,& & & & & \\
$^{162}$Nd, $^{166}$Sm & & & & & \\
$^{236}$Bk, $^{240}$Es  & J. Konki & 10/4/2016 &\refcite{2017Kon01} & Jyv\"askyl\"a & FE \\
$^{81}$Mo, $^{82}$Mo, $^{85}$Ru, $^{86}$Ru  & H. Suzuki & 10/19/2016 &\refcite{2017Suz01} & RIKEN & PF \\
$^{223}$Np  & M. D. Sun & 11/21/2016 &\refcite{2017Sun01} & Lanzhou & FE \\
$^{77}$Zr, $^{72}$Rb  & H. Suzuki & 4/3/2017 &\refcite{2017Suz02} & RIKEN & PF \\
$^{17}$Na  & K. W. Brown & 4/6/2017 &\refcite{2017Bro01} & Michigan State & SB \\
\botrule
\end{tabular}}
\end{table}

The discovery of six neutron-deficient nuclides was described by Sumikama et al. in the paper ``Observation of new neutron-rich Mn, Fe, Co, Ni, and Cu isotopes in the vicinity of $^{78}$Ni'' \cite{2017Sum01}. A 3-mm-thick beryllium target was irradiated with a 345 MeV/nucleon $^{238}$U from the RIKEN Radioactive Isotope Beam Factory (RIBF). Fission fragments were identified behind the BigRIPS separator and the ZeroDegree spectrometer: ``The particle-identification plot for the in-flight fission fragments highlights the first observation of eight new isotopes: $^{73}$Mn, $^{76}$Fe, $^{77,78}$Co, $^{80,81,82}$Ni, and $^{83}$Cu.'' Sumikama et al. are credited only with the discovery of six nuclides because $^{77}$Co and $^{80}$Ni had previously been reported by Xu et al.\cite{2014Xu01} as acknowledged in the paper.

Wu et al. discovered another 18 new nuclides in ``94 $\beta$-Decay Half-Lives of Neutron-Rich $_{55}$Cs to $_{67}$Ho: Experimental Feedback and Evaluation of the r-Process Rare-Earth Peak Formation'' \cite{2017Wu01}. A 345 MeV/nucleon $^{238}$U beam from RIBF was used to produce fission fragments which were separated with the BigRIPS separator and the ZeroDegree spectrometer. The nuclides were identified with the TOF-B$\rho$-$\Delta$E method and their $\beta$-decay half-lives were measured with the Wide range Active Silicon-Strip Stopper Array for Beta and ion detection (WAS3ABi). The most exotic isotope measured for each element was indicated in a particle identification plot and the extracted half-lives were listed in a table as supplemental material. See section \ref{Changes2017} for a discussion of the cesium isotopes $^{149-151}$Cs.

$^{240}$Es and $^{236}$Bk were first reported in the paper ``Towards saturation of the electron-capture delayed fission probability: The new isotopes $^{240}$Es and $^{236}$Bk'' by Konki et al.\cite{2017Kon01}. A $^{34}$S beam was accelerated to 174 and 178 MeV with the K-130 cyclotron of the Accelerator Laboratory of the Department of Physics, University of Jyv\"askyl\"a, Finland and bombarded $^{209}$BiO$_2$ foils. Evaporation residues were separated with the gas-filled recoil separator RITU and identified with the GREAT focal plane spectrometer. ``Four chains with E$_{\alpha_1}$ = 8.09 MeV and one with E$_{\alpha_1}$ = 8.19 MeV were observed. These chains were attributed to originate from the $\alpha$ decay of $^{240}$Es, which then proceeds to $^{236}$Cm through the EC decay of $^{236}$Bk.'' The half-lives for $^{240}$Es and $^{236}$Bk were extracted to be 6(2)~s and 22$^{+13}_{-6}$ s, respectively.

Suzuki et al. reported the discovery of the four neutron-deficient nuclides $^{81,82}$Mo and $^{85,86}$Ru in ``Discovery of new isotopes $^{81,82}$Mo and $^{85,86}$Ru and a determination of the particle instability of $^{103}$Sb'' \cite{2017Suz01}. A 345 MeV/nucleon $^{124}$Xe from the RIKEN Nishina Center RI Beam Factory bombarded a 4.03-mm-thick beryllium target and the fragmentation products were separated with the BigRIPS separator. ``As shown in the figure, we have clearly identified four new isotopes $^{81}$Mo, $^{82}$Mo, $^{85}$Ru, and $^{86}$Ru. The numbers of observed events were 1, 6, 1, and 35, respectively.'' These results had been published a few years earlier in a conference proceeding \cite{2013Suz01}.

$^{223}$Np was discovered in ``New short-lived isotope $^{223}$Np and the absence of the Z=92 subshell closure near N=126'' by Sun et al.\cite{2017Sun01}. $^{40}$Ar was accelerated with the Sector-Focusing Cyclotron (SFC) of the Heavy Ion Research Facility in Lanzhou (HIRFL) to 188 MeV and bombarded an enriched $^{187}$Re target. $^{223}$Np was formed in the fusion-evaporation reaction $^{187}$Re($^{40}$Ar,4n) and implanted in a double-sided silicon strip detector located after the recoil separator SHANS. ``The half-life of $^{223}$Np was determined to be 2.15($^{100}_{52}$)$\mu $s by averaging the time differences between $^{223}$Np implantations and decays...'' 

The first identification of $^{77}$Zr and $^{72}$Rb was reported by Suzuki et al. in the paper ``Discovery of $^{72}$Rb: A Nuclear Sandbank Beyond the Proton Drip Line'' \cite{2017Suz02}. A 345 MeV/nucleon $^{124}$Xe beam from RIBF impinged on a 740 mg/cm$^2$ thick beryllium target. In-fight fragments were separated by BigRIPS and the ZeroDegree  spectrometer and implanted in the active silicon stopper WAS3ABi. ``For the half-life of the new isotope $^{72}$Rb, we obtained 103(22) ns, from an upper limit of 124 ns [...], and a lower limit of 81 ns, [...] Tentative evidence for the existence of $^{77}$Zr was also observed with one count.''

Brown  et al. discovered the proton-unbound nuclide $^{17}$Na as described in ``Proton-decaying states in light nuclei and the first observation of $^{17}$Na'' \cite{2017Bro01}. A secondary $^{17}$Ne beam was produced at the Coupled Cyclotron Facility of the National Superconducting Cyclotron Laboratory at Michigan State University from a 150 MeV/nucleon primary $^{20}$Ne beam and was fragmented on a beryllium target. Charged particles from the particle-unbound fragments were detected in the High Resolution Array (HiRA). $^{17}$Na could be produced in a charge-exchange reaction. ``The decay-energy spectrum for $^{17}$Na $\rightarrow$ 3p + $^{14}$O is shown in the figure. There is a peak in the spectrum located at E$_T$ = 4.85(6) MeV that sits on top of a background.''

\section{Changes in  2017}
\label{Changes2017}

Originally, the discoveries of $^{149-152}$Cs were attributed \cite{2012May01,2016Tho01} to a 1987 paper by Ravn \cite{1987Rav01}, however, Ravn did not extract any quantities of these nuclides nor did he present any evidence for its direct observation. The half-lives of $^{149}$Cs and $^{150}$Cs were reported in the 1999 Ph.D thesis by K\"oster\cite{1999Kos01} but were not published in a refereed journal. As mentioned above in section \ref{New2017}, $^{149}$Cs, $^{149}$Cs, and $^{150}$Cs  were observed by Wu et al. in 2017\cite{2017Wu01} while $^{152}$Cs remains unobserved.
It should be mentioned that the half-lives of $^{149}$Cs and $^{150}$Cs were also finally published in 2017\cite{2017Lic01}, however, the paper was submitted on 11/15/2016, about seven weeks (9/29/2016) after Wu et al. had submitted their direct observation of these isotopes.

In 1995 Rykaczewski et al. reported the discovery of six neutron-deficient nuclides near $^{100}$Sn\cite{1995Ryk01}. Two of the new nuclides ($^{89}$Rh and $^{103}$Sb) were predicted to be proton unbound and thus their observation was somewhat unexpected. Recent experiments did not confirm the observation of these nuclides so that these assignments had to be retracted.  In a projectile fragmentation reaction at RIKEN, Celikovic et al. did not observe any events of $^{89}$Rh and deduced on upper half-life limit of 120 ns\cite{2016Cel01}. Similarly, in the experiment discovering $^{81,82}$Mo and $^{85,86}$Ru Suzuki et al. did not observe any events of $^{103}$Sb extracting an upper half-life limit of 46 ns\cite{2017Suz01}. With these short half-lives Rykaczewski et al. could not have observed these isotopes as the flight path in their experiment was 118 m long corresponding to a time of flight of about 1.5$\mu$s.

\section{Status at the end of 2017}

The 34 new discoveries in 2017 correspond to the largest number of nuclides discovered per year since 2012 and the fourth largest over the last twenty years. Accounting for the six retractions, they increased the total number of observed isotopes to 3252. They were reported by 911 different first authors in 1543 papers and a total of 3717 different coauthors.

\begin{figure}[pt] 
\centerline{\psfig{file=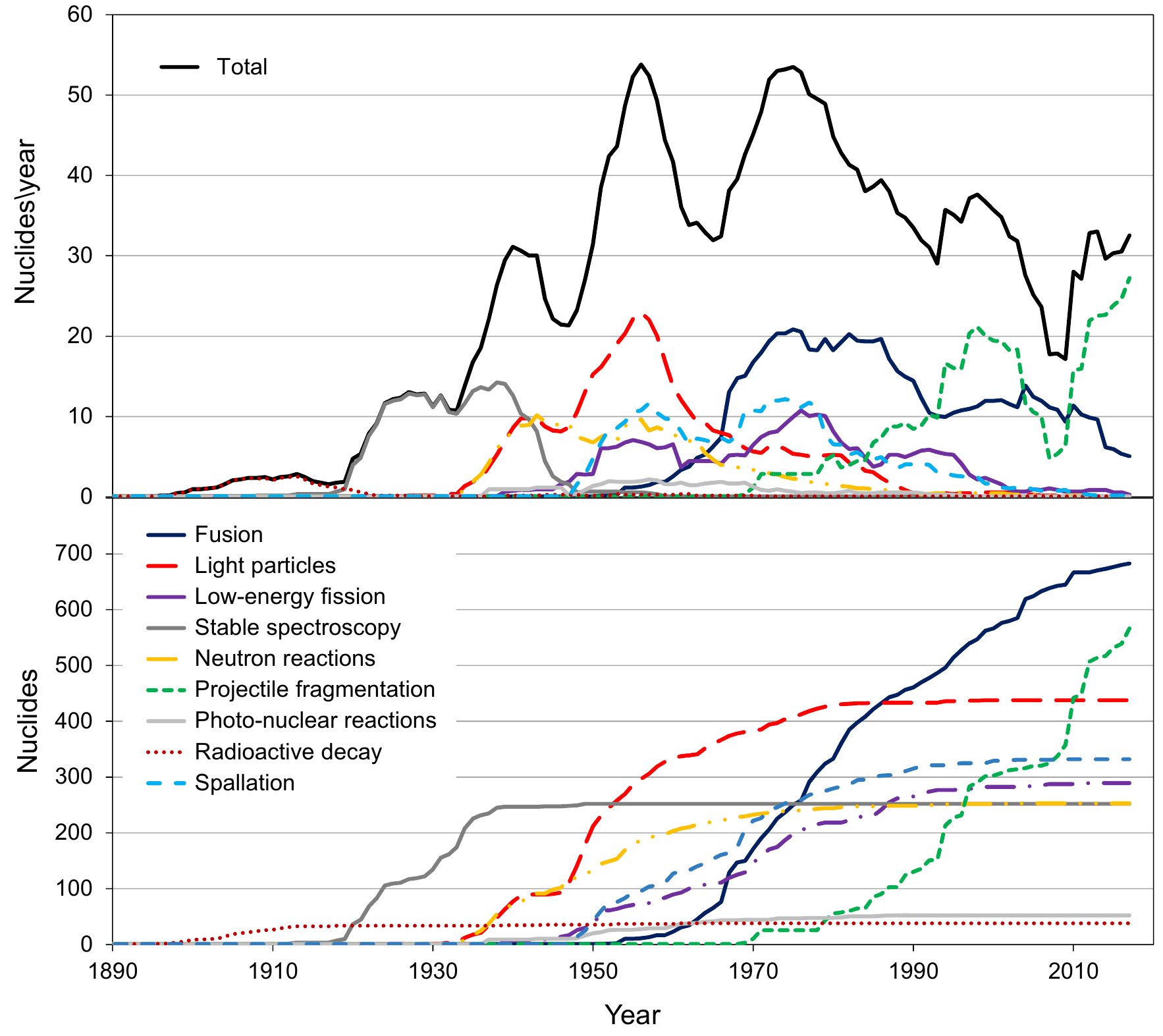,width=12.4cm}}
\caption{Discovery of nuclides as a function of year. The top figure shows the 10-year running average of the number of nuclides discovered per year while the bottom figure shows the cumulative number. The different colored lines correspond to the different methods used to produce the nuclides as shown in the bottom figure. The top figure also shows the total number of nuclides (black line). This figure was adapted from Ref. 5 to include the data from 2017\label{f:timeline}. }
\vspace*{-0.1cm}
\end{figure}

With the 30 new nuclides observed in 2017, a total of 138 nuclides have been discovered at RIKEN. This corresponds to the 5$^{th}$ largest number for any laboratory, only behind Berkeley (639), GSI (438), Dubna (221), and Cambridge (218). After the retraction of the four cesium isotopes, CERN shares the 6$^{th}$ rank with Argonne with 111 nuclides. GANIL (84), Oak Ridge (79), and Michigan State (74) complete the list of top ten laboratories.

The order for the top countries remained the same: USA (1328), Germany (559), UK (299), Russia (249), France (220), Japan (150), Switzerland (127), Canada (66), Sweden (59) and Finland (40).  Further statistics can be found on the discovery project website \cite{2011Tho03}.

Figure \ref{f:timeline} shows the current status of the evolution of the nuclide discoveries for the main means of production as labeled in the figure. The figure was adapted  from the 2014 review\cite{2014Tho01} and was extended to include all isotopes discovered until the end of 2017. The top part of the figure shows the ten-year average of the number of nuclides discovered per year while the bottom panel shows the integral number of nuclides discovered. The top figure includes the total of nuclides while the bottom figure only shows the contributions of the individual production methods.

Although the ten-year average rate for nuclides produced in projectile fragmentation (which also includes in-flight projectile fission and secondary beams) reached yet another all-time high of 27.2, overall there are still more nuclides produced in fusion-evaporation reactions (768) than in projectile fragmentation reactions (637). In the last 20 years these two mechanisms accounted for 96\% of all discovered nuclides.

\section{Discoveries not yet published in refereed journals}

Nuclides which have so far only been presented in conference proceedings or internal reports are listed in 
Table \ref{reports}. It is significantly shorter (60) than last year's list (90)\cite{2017Tho01} although it contains an additional ten new light neutron-rich nuclides from $^{39}$Na through $^{62}$Sc produced in projectile fragmentation reactions at RIBF\cite{2016Ahn01,2017Tar01}.

Eighteen of the nuclides listed in last year's table were discovered in 2017 by Wu\cite{2017Wu01} and Suzuki\cite{2017Suz01}.  Another 20 nuclides are included in two papers which will be published in 2018 but were already available online at the end of 2017. Fukuda et al. published the new observation of $^{163}$Nd, $^{164}$Pm, $^{165}$Pm, $^{167}$Sm, $^{169}$Eu, $^{171}$Gd, $^{173}$Tb, $^{174}$Tb, $^{175}$Dy, $^{176}$Dy, $^{177}$Ho, and $^{179}$Er\cite{2018Fuk01} and Shimizu et al. reported the new nuclides $^{122}$Tc, $^{125}$Ru, $^{130}$Pd, $^{131}$Pd, $^{140}$Sn, $^{142}$Sb, $^{146}$I, and $^{157}$La\cite{2018Shi01}. The detailed description of these papers will be included in the 2018 update. Also, a recent paper by Kaji et al. could not yet unambiguously identify $^{280}$Ds\cite{2017Kaj01}.

\begin{table}[t]
\tbl{Nuclides only reported in proceedings or internal reports until the end of 2017. The nuclide, first author, reference and year of proceeding or report are listed. \label{reports}}
{\begin{tabular}{@{}llrr@{}} \toprule
\parbox[t]{6.8cm}{\raggedright Nuclide(s) } & \parbox[t]{2.3cm}{\raggedright First Author} & Ref. & Year \\ \colrule

$^{20}$B		&	 F. M. Marqu\'es 	&	\refcite{2015Mar01}	&	2015	 \\ 
$^{21}$C		&	 S. Leblond 	&	\refcite{2015Leb01,2015Leb02}	&	2015	 \\ 
			&	 N. A. Orr 	&	\refcite{2016Orr01}	&	2016	 \\ 
$^{39}$Na	& D. S. Ahn 	&	\refcite{2016Ahn01}	&	2016	 \\ 
			& O. B. Tarasov 	&	\refcite{2017Tar01}	&	2017	 \\ 
$^{47}$P, $^{49}$S, $^{52}$Cl, $^{54}$Ar, $^{57}$K, $^{59}$K, $^{59}$Ca, $^{60}$Ca, $^{62}$Sc	& O. B. Tarasov 	&	\refcite{2017Tar01}	&	2017	 \\ 
$^{86}$Zn, $^{88}$Ga, $^{89}$Ga, $^{91}$Ge, $^{93}$As, $^{94}$As, $^{96}$Se, $^{97}$Se	&	 Y. Shimizu 	&	\refcite{2015Shi01}	&	2015	 \\ 
$^{99}$Br, $^{100}$Br & & & \\
$^{98}$Sn, $^{104}$Te	&	  I. Celikovic 	&	\refcite{2013Cel01}	&	2013	 \\ 
$^{155}$Ba, $^{159}$Ce, $^{161}$Pr, $^{164}$Nd, $^{166}$Pm, $^{168}$Sm,  	&	 N. Fukuda 	&	\refcite{2015Fuk01}	&	2015	 \\
$^{170}$Eu, $^{172}$Gd, $^{173}$Gd, $^{175}$Tb, $^{177}$Dy, $^{178}$Ho, $^{179}$Ho,& & & \\
 $^{180}$Er, $^{181}$Er, $^{182}$Tm, $^{183}$Tm & & & \\
$^{126}$Nd, $^{136}$Gd, $^{138}$Tb, $^{143}$Ho$^a$, $^{150}$Yb, $^{153}$Hf	&	 G. A. Souliotis 	&	\refcite{2000Sou01}	&	2000	 \\
	$^{143}$Er, $^{144}$Tm	&	 R. Grzywacz 	&	\refcite{2005Grz01}	&	2005	 \\
	& K. Rykaczewski & \refcite{2005Ryk01} & 2005 \\
	& C. R. Bingham & \refcite{2005Bin01} & 2005 \\
$^{230}$At, $^{232}$Rn	&	 J. Benlliure 	&	\refcite{2010Ben02}	&	2010	 \\
							&	  	&	\refcite{2015Ben01}	&	2015	 \\
$^{235}$Cm	&	 J. Khuyagbaatar 	&	\refcite{2007Khu01}	&	2007	 \\
$^{252}$Bk, $^{253}$Bk	&	 S. A. Kreek 	&	\refcite{1992Kre01}	&	1992	 \\
$^{262}$No 	&	 R. W. Lougheed 	&	\refcite{1988Lou01},\refcite{1989Lou01}	&	1988/89	 \\
	&	 E. K. Hulet 	&	\refcite{1989Hul01}	&	1989	 \\
$^{261}$Lr, $^{262}$Lr	&	 R. W. Lougheed 	&	\refcite{1987Lou01}	&	1987	 \\
	&	 E. K. Hulet 	&	\refcite{1989Hul01}	&	1989	 \\
	&	 R. A. Henderson 	&	\refcite{1991Hen01}	&	1991	 \\
$^{255}$Db	&	 G. N. Flerov	&	\refcite{1976Fle01}	&	1976	 \\
		& A.-P. Lepp\"anen	& \refcite{2005Lep01} & 2005 \\
$^{280}$Ds	&	 K. Morita	&	\refcite{2015Mor01}	&	2014	 \\
\botrule
\vspace*{-0.2cm} & & & \\
$^a$ also published in ref. \refcite{2003Sew02} \\
\end{tabular}}
\end{table}

In addition to the nuclides listed in table \ref{reports}, seven new nuclides had been presented at the 4$^{th}$ Joint Meeting of the APS Division of Nuclear Physics and the Physical Society of Japan in 2014 \cite{2014Shi01}. Two of these nuclides ($^{138}$In and $^{143}$Sb) remain unpublished as Wu et al. reported the observation of $^{154}$Ba\cite{2017Wu01} and Shimizu et al. claimed the discovery of $^{116}$Nb, $^{145}$Te, $^{147}$I, and $^{149}$Xe\cite{2018Shi01}.

\section{Summary}

The 34 new isotopes discovered in 2017 represent a significant increase relative to the previous four years when on average only 13 nuclides were discovered per year. This positive trend promises to be continued next year as at this point the discovery of 33 new nuclides are already scheduled to be published in 2018. The articles describing these observations became available online at the end of 2017. It is expected that the analysis of a large fraction of the not-yet-published nuclides observed at RIKEN will be completed soon and should be ready for publication during the next year.

\section*{Acknowledgements}

Support of the National Science Foundation under grant No. PHY15-65546 is gratefully acknowledged.


\bibliographystyle{ws-ijmpe}
\bibliography{update}

\end{document}